\documentclass[prl,showpacs,twocolumn,superscriptaddress]{revtex4}
\usepackage{amsmath}
\usepackage{amsfonts}
\usepackage{epsfig}
\usepackage{color}

\begin{document}
\title{A topological charge selection rule for phase singularities}

\author{M. Zacar\'es}
\affiliation{Institut de Matem\`atica Pura i Aplicada. Universitat Polit\`ecnica de Val\`encia, Val\`encia, Spain}
\author{M.A. Garc\'{\i}a-March}
\affiliation{Institut de Matem\`atica Pura i Aplicada. Universitat Polit\`ecnica de Val\`encia, Val\`encia, Spain}
\author{J. Vijande}
\affiliation{Departamento de F\'{\i}sica Te\'{o}rica, Universidad de Valencia (UV)  and IFIC (UV-CSIC), Burjassot, Spain.}
\author{A. Ferrando}
\affiliation{Departament d'\'Optica, Universitat de Val\`encia, Val\`encia, Spain}
\author{E. Merino}
\affiliation{Departamento de F\'{\i}sica Te\'{o}rica, Universidad de Valencia (UV) and IFIC (UV-CSIC), Burjassot, Spain.}
\date\today

\begin{abstract}
We present an study of the dynamics and decay pattern of phase singularities
due to the action of a system with a discrete rotational symmetry of finite order.
A topological charge conservation rule is identified.
The role played by the underlying symmetry is emphasized. 
An effective model describing the short range dynamics of the vortex clusters has 
been designed. A method to engineer 
any desired configuration of clusters of phase singularities is proposed.
Its flexibility to create and control clusters of vortices is discussed.

\end{abstract}
\pacs{02.40.Xx,42.65.Sf,03.75.Lm,42.50.Tx}
\maketitle

With the coming of the new century the study and understanding of the formation and dynamics of phase singularities in 
nature has become one of the major milestones in modern physics. The relevance of one particular kind of phase 
singularities, the vortices, has been made evident from the variety of scientific fields where they play a major role: 
nuclear physics, condensed-matter physics, optics, superfluidity, cosmology...~\cite{Sch89}

This would have not been possible without the acknowledgement of the important role played by 
symmetry principles in our understanding of nature. With the development of physics in the 20th century, symmetry 
considerations evolved from a {\it passive role} in which symmetry was a property of the 
interactions to an {\it active role} in which symmetry serves to determine the interactions and the conservation laws themselves, as 
C.N. Yang stated {\it ``symmetry dictates interaction``}~\cite{Yan80}. 

Although the following results and discussion can be applied to any kind of phase singularity we choose one particular
system to work without any loss of generality, the optical vortices. In optics, 
or in any kind of general theory of wave mechanics, vortices are solutions of the wave equations characterized by a 
single phase singularity, i. e., a  screw phase dislocations where the amplitude vanishes~\cite{Nye74}. The phase around the 
singularity has an integer number of windings. For symmetric systems, this number, usually denoted as $v$, is a 
conserved quantity and governs the interactions between vortices as if they were endowed with electrostatic charges. 
Thus, $v$ is referred to in the literature as vorticity, topological charge, or winding number. These systems have been widely discussed in the field of nonlinear optics, both in homogeneous~\cite{Seg98} and discrete symmetry media~\cite{Efr02}. 

Phase singularities in general, and vortex solitons
in particular, behave and interact with each other in the same way as particles do ~\cite{Zab65}. In fact, skyrmions, solitons whose 
topological charge has been identified with the baryon number, have been widely used in nuclear 
physics to obtain an effective description of nuclei~\cite{Sky62}. 
Symmetries and their corresponding conservation laws play a crucial role in the description of 
particles, interactions and  decay patterns, being the Standard Model (SM) of electroweak interactions 
a textbook example~\cite{Sal69}. 
Therefore, pushing even forward the 
phase singularity-particle analogy, one may wonder whether it is possible to characterize the 
dynamics of the phase singularities as particle-like systems by using symmetry arguments. 

In this letter we take advantage of symmetry principles and group theory arguments to address the phase singularities  dynamics and decay pattern, i.e., the formation of clusters of phase singularities as a result of the break-up of the initial one, 
in the vicinity of an interface between an homogeneous medium characterized by an $O(2)$ symmetry and a discrete $C_n$ one.

An optical vortex within a medium characterized by a continuous $O(2)$ 
symmetry presents well-defined angular momentum $\ell$. Such a system being an ideal candidate 
for new technological developments~\cite{FraAr08}.
In systems such as optical media with discrete symmetry~\cite{Efr02} or Bose-Einstein condensates 
in 2D periodic traps~\cite{Bai03}, the $O(2)$ symmetry is replaced by a discrete-point symmetry, $C_{n}$, and the angular
momentum becomes ill-defined. Nevertheless, although 
these solutions cannot have well-defined angular momentum any longer, certainly 
all of them present neat phase dislocations that can be characterized by
an integer value. In this case one can define an integer quantity $m$, the  angular pseudomomentum,
which is conserved during propagation \cite{Fer05}.
This behaviour is analog to the one of particles crossing into a discrete periodic medium, usually electrons entering into a crystalline system. In these media the linear momentum of the particle is no longer a conserved quantity, being replaced 
on this role by the Bloch wavevector~\cite{Ash76}.

In the case of homogeneous media, angular momentum and vorticity coincide, $\ell=v$. However we shall see that
this is no longer true whenever the vortex moves into a medium characterized by a $C_n$ discrete symmetry.
To discuss its dynamics we will take advantage of discrete group theory arguments in the same way it was done in 
Refs.~\cite{Fer05}. As initial condition we consider the propagation inside an $O(2)$ medium 
of a vortex with angular momentum $\ell$ given by $\phi_{\ell}(r,\theta,z)=e^{\imath\ell\theta}f_{\ell}(r,z)$, 
where $f_{\ell}(r,z)$ describes the radial dependence of the vortex amplitude. The equation of motion that governs 
the dynamics of the field in both media is the standard nonlinear Schr\"odinger equation given by
\begin{equation}
\label{Eq3}
\imath\frac{\partial\phi_{\ell}}{\partial z}=[-\nabla_{2D}^2+V_L(x,y,z)+V_{NL}(|\phi_{\ell}|)]\phi_{\ell}
\end{equation}
where $V_L(x,y,z)=V_{C_n}(r,\theta)H(z)+V(r)_{O(2)}[1-H(z)]$ characterize the media before and after the
interface, $H(x)$ is the Heaviside step function and $V_{O(2)}(r)$ 
and $V_{C_n}(r,\theta)$ are rotationally invariant potentials, $O(2)$ the former and $C_n$ the latter. 
$V_{NL}(|\phi|)$ is a nonlinear term depending on the field modulus. 

In the linear regimen, given an initial condition and knowing the eigenstates (eigenmodes), $\psi_p(x,y)$, and eigenvalues (propagation constant), $\mu_p$, of the system, the field $\phi_{\ell}$ is completely determined for any value of $z$ as $\phi_{\ell}(r,\theta,z)=\sum_pe^{\imath\mu_pz}c_p(0)\psi_p(r,\theta)$. In the nonlinear case this is no longer true. However, one can perform the same expansion in an infinitesimal element $dz$ in which the nonlinear term varies slowly and therefore the conclusions holds. The $c_p(0)$ coefficients can be obtained through the projection of the vortex over the eigenmodes of the system. Thus, given the system eigenmodes and their projections over the interface the vortex evolution is completely determined. 

Group theory allows us to go one step further and extract more information about these coefficients. If the linear and nonlinear potentials are invariant under the $C_n$ group, the eigenmodes of the system can be classified according to the different group representation. By using this, the field over the interface can be rewritten as $\phi_{\ell}(r,\theta,0)=\sum_{m,\overline{m}}C_{m\overline{m}}^{\ell}e^{\imath m\theta}U_{m\overline{m}}(r,\theta)$, where $\overline{m}$ numerate the modes with  angular pseudomomentum $m$ in each representation.
Since $\phi_{\ell}(r,\theta,0)$ and $U_{m\overline{m}}(r,\theta)$ belong to representations of $O(2)$ and $C_n$ respectively, they both transform properly under a discrete rotation of order $n$. Thus, by performing the change of variable $\theta\to\theta+2\pi/n$ one arrives to the symmetry relation $C_{m\overline{m}}^{\ell}=e^{\imath 2\pi(\ell-m)/n}C_{m\overline{m}}^{\ell}$. Then, the $C_{m\overline{m}}^{\ell}$ coefficients are zero unless the condition 
\begin{equation}
\label{Eq1}
\ell=m+k_0n\,\,\,(k_0\in\mathbb{Z})
\end{equation}
is fulfilled. Even more, in Refs. \cite{Fer05} it was proved that the values of $m$ are further constrained by the 
order of the point-symmetry group $C_n$, being $|m|\leq n/2$ (even n) or $|m|\leq (n-1)/2 $ (n odd).
 
Eq.(\ref{Eq1}) also implies that once the angular momentum and the symmetry order are fixed (initial conditions of our problem), 
the vortex would only be projected over modes belonging to the $m$-th representation of the symmetry group, 
i.e. the dynamics of the vortex will be described in terms of eigenmodes characterized by an angular pseudomomentum  $m$. 
This implies that the field over the interface can be expressed as a linear combination of the form
$\phi_{mk_0}(r,\theta,0)=\sum_{\overline{m}}C^{k_0}_{m\overline{m}}e^{\imath m\theta}U_{m\overline{m}}(r,\theta)$.
$U_{m\overline{m}}(r,\theta)$ are periodic functions in the
angular variable and therefore they admit a Fourier expansion. Introducing this into the definition of 
$\phi_{mk_0}(r,\theta,0)$ one arrives to $\phi_{mk_0}(r,\theta,0)=\sum_{k,\overline{m}}C_{m\overline{m}}^{k_0}e^{\imath (m+kn)\theta}U^k_{m\overline{m}}(r)$. 
Knowing the expression of the coefficients over the interface and summing over the subindex $\overline{m}$, the 
expression of the field for any value of $z$ is given by
\begin{equation}
\label{Eq2}
\phi_{mk_0}(r,\theta,z)=\sum_{k}a_m^k(r,z)e^{\imath (m+kn)\theta}
\end{equation}
where $a_m^k(r,z)=\sum_{\overline{m}}C_{m\overline{m}}^{k_0}e^{-\imath \mu_{m\overline{m}}z}U^k_{m\overline{m}}(r)$.
Eq.(\ref{Eq2}) can be understood as a linear combination of vortex functions whose charges are completely determined by the angular pseudomomentum $m$ and the order of the group $n$.  The dynamics of each one being determined by the coefficients $ a_m^k(r,z)$. This expression is valid in the linear and in the nonlinear case. For the latter, it is important to emphasize that $V_{NL}(|\phi_{\ell}|)$ depends only on the modulus of the field, and therefore the eigenmodes have well defined symmetry properties under the $C_n$ discrete group. This makes that Eq.(\ref{Eq2}) will be valid within an infinitesimal interval $dz$. If we consider the propagation on the interval $[0,z]$ as a sum of subintervals $[z_i,z_i+dz]$ and we apply the aforementioned arguments to each one of them, it can proved that Eq.(\ref{Eq2}) would also be valid in the nonlinear case.

The evolution of the radial coefficients is obtained by introducing Eq.(\ref{Eq2}) into Eq.(\ref{Eq3}) and integrating out the angular part,
\begin{eqnarray}
\label{Eq4}
\imath\frac{\partial a_m^q}{\partial z}&=&\left[-\frac{\partial^2}{\partial r^2}-\frac{1}{r}\frac{\partial}{\partial r}+
\frac{(m+qn)^2}{r^2}\right]\nonumber\\&&+\sum_k\overline{V}_{q-k}a_m^k+\sum_kM_{q-k}a_m^k
\end{eqnarray}
where $\overline{V}_{q-k}$ is given by the Fourier expansion of $V_{C_n}(r,\theta)$. 
The functional form of $M_{q-k}$ will depend on the kind of nonlinearity considered. For Kerr nonlinearities it 
will take the form $M_{q-k}=\gamma(z)\sum_pa^*_{k-q+p}a_p$. The dynamics described by Eq. (\ref{Eq4}) is completely 
equivalent to the one given by the nonlinear Scr\"odinger equation. However, in this case 
we have reduced the dimensionality of the problem, obtaining a description of the nonlinear dynamics in terms only 
of the radial variables.

To provide a physical example we consider an optical interface separating two 2D dielectric media with Kerr nonlinearity, 
these two media being a homogeneous medium and a 2D square optical lattice. This system is equivalent to a 2D BEC 
in which a periodic
potential is abruptly switched on. They constitute an $O(2)-C_4$ interface described by Eqs.~(\ref{Eq3}) and~(\ref{Eq4}) with
$V_{NL}=|\phi_{\ell}|^2$, $V(r)_{O(2)}=V_0$, and
$V_{C_n}(x,y)=V_1\sum_{j=0}^{n-1}\text{Exp}\left\{\frac{-(x-x_j)^2+(y-y_j)^2}{2\omega^2}\right\}$, where $(x_j,y_j)=d(\cos(2\pi j/n,\sin(2\pi j/n))$ ($d$ stands for the distance between gaussians, and $V_1=-1$). 

\begin{figure}[tb]
\centering
\begin{tabular}{ccc}
(a)&(b)&(c)\tabularnewline
\includegraphics[scale=0.48]{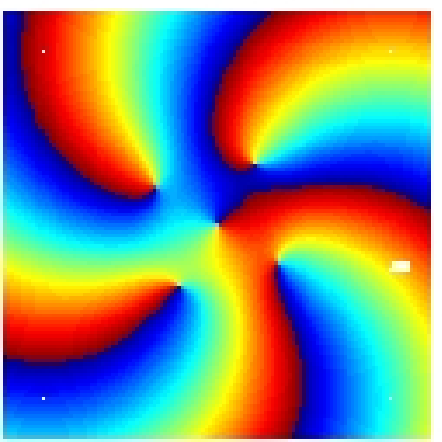}&
\includegraphics[scale=0.21]{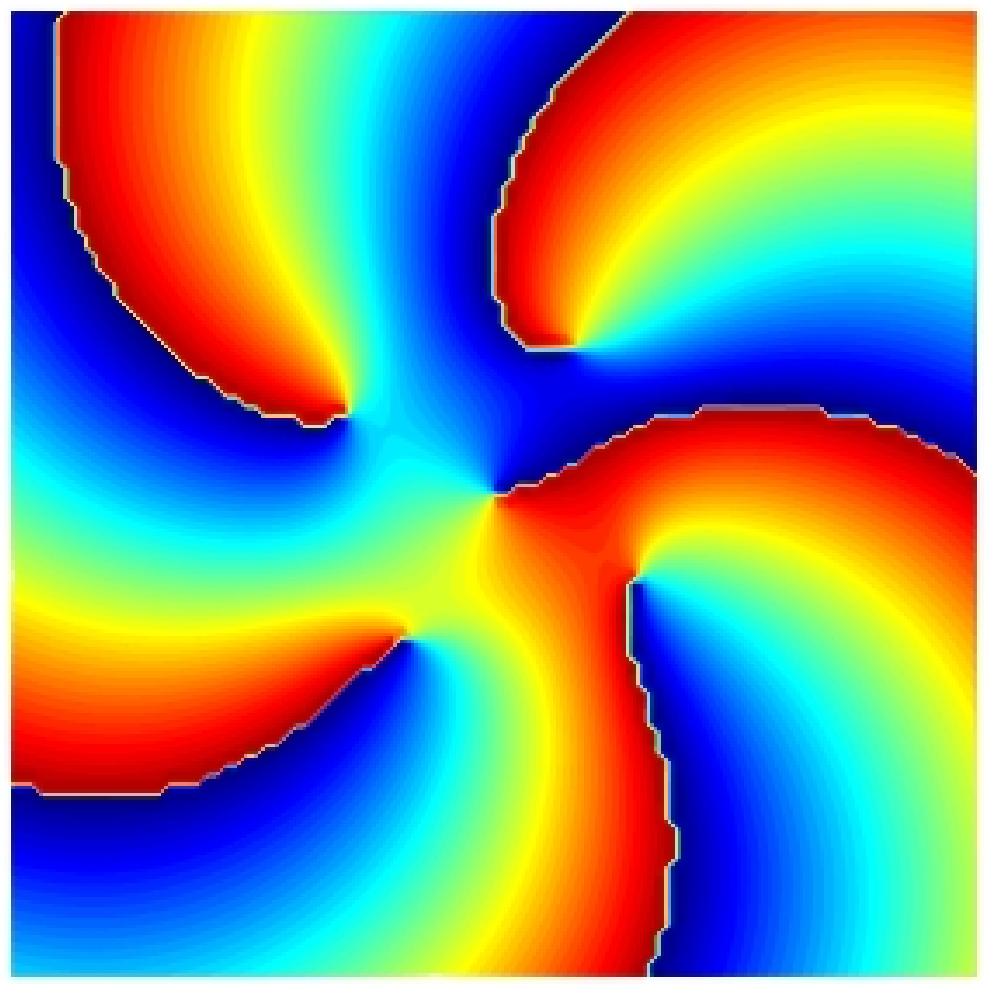}&
\includegraphics[scale=0.29]{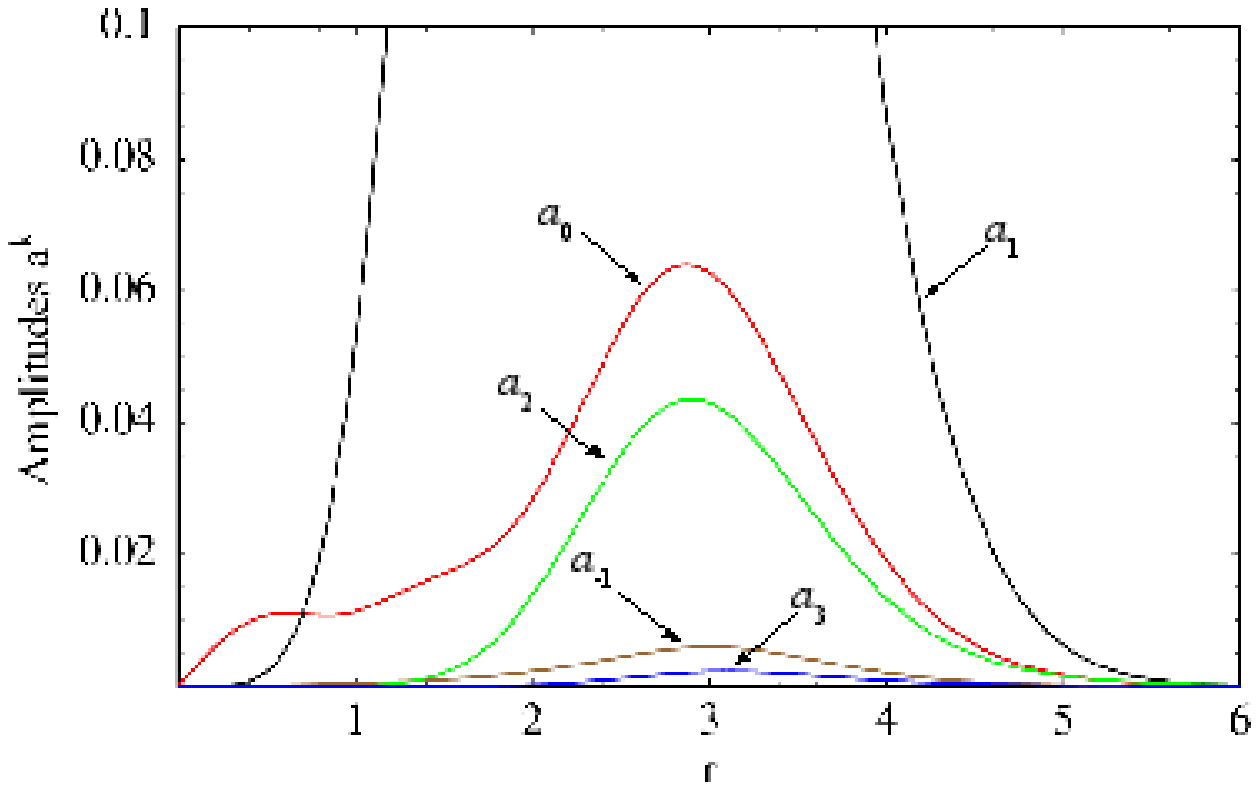}\tabularnewline
\end{tabular}
\caption{Phase calculated for $z=0.30$ (a) using Eq.~(\ref{Eq3}) and (b) Eq.~(\ref{Eq4}). The corresponding angular components $a_m^q(r,z=0.30)$  are given in (c). }
\label{f2}
\end{figure}

We consider an initial vortex with charge $\ell=5$ crossing the interface into the $C_4$ symmetric medium. Thus, by applying Eq.~(\ref{Eq1}) one obtains $k_0=1$. The initial condition in the interface will be given by $a_m^k(r,0)=0$ if $k\neq k_0$ and $a_m^k(r,0)=f_{\ell}(r,0)$ if $k=k_0$, therefore the only nonzero amplitude in the interface will be $a_m^1$.  Since we are interested in the short distances limit ($z\to 0,r\to0)$, Eq.~(\ref{Eq2}) can be safely truncated in the vicinity of the symmetry axis at second order. Being consistent with this approximation we have considered a potential in which the Fourier expansion can be truncated in the same way, $V(r,\theta)\approx \overline{V}_{-1}(r)e^{-\imath 4\theta}+\overline{V}_{0}(r)+\overline{V}_{1}(r)e^{\imath 4\theta}$ where $\overline{V}_0>\overline{V}_{\pm1}$. 
We show in Fig.~\ref{f2}c) that only a reduced set of angular amplitudes, $a_m^0$, $a_m^1$, and $a_m^2$, play a significant role in the description of the dynamics of the system in the short distances limit. In Fig.~\ref{f2} we present the phase $\varphi$ of the complex solution $\phi=|\phi|e^{i\varphi}$ obtained numerically solving Eq.~(\ref{Eq4}) and the complete non-linear Schr\"odinger equation Eq.~(\ref{Eq3}), to show that both simulations provide similar results. From these pictures it can be observed one singularity lying in the symmetry axis with charge +1, note that according to Eq.~(\ref{Eq2}) $m=(\ell=5)-(k_0n=4)=+1$, together with four 
singularities with charge +1 symmetrically distributed around it. The same breaking pattern is observed for other  simulations with different $\ell$ and $n$ provided that $k_0=1$, i.e., one static singularity in the center with charge $m$ and clusters of $n$ singularities with charge +1 moving away from the symmetry axis, see EPAPS document No. [EPAPS-01.gif] for some examples. Let us establish the physical meaning of the angular coefficients $a_m^k$. As shown in Fig.~\ref{f2}, for $r\rightarrow 0$ and $z\rightarrow 0$  only $a_m^0$ and $a_m^1$ must be considered. Therefore, in this limit, Eq.~(\ref{Eq2}) can be written as $\phi(r,\theta,z)\approx a_m^0(r,z)e^{im\theta}+a_m^1(r,z)e^{i(m+n)\theta}$. 
Note that, if $a_m^0\propto r^m$ and $a_m^1\propto r^{m+n}$ this equation is transformed in  $\phi(r,\theta,z)\approx r^m e^{im\theta}+ r^{m+n}e^{i(m+n\theta)}$, which can be written as $\phi(r,\theta,z)\approx \omega^m(a+b \omega^{n})$ with $\omega\in\mathbb{C}$, $a,b\in\mathbb{R}$, and then it correctly describes the clusterization pattern observed numerically.  The behaviour $a_m^k\propto r^{|m+kn|}$ is not that surprising since it has been demonstrated for stationary solutions, i.e. solutions $\phi(r,\theta,z)=e^{i\mu z}\psi(r,\theta)$~\cite{GaMa08}. It has been checked numerically that $a_m^0$ and $a_m^1$ present this behaviour for $r\rightarrow0$.  

\begin{figure}[t]
\centering
\includegraphics[scale=0.5]{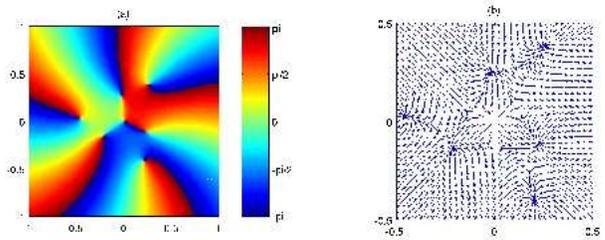}
\caption{Phase (a) and dual of the flux field (b) for an initial vortex with angular momentum $\ell=5$ after evolving into a $C_3$ symmetric medium.}
\label{f3}
\end{figure}

Once the importance of $\ell$ and $n$ has been established, let us show the pivotal role played by  $k_0$ in the mechanism of the formation of clusters of vortices. To do so those solutions where $k_0>1$ are of particular interest. 
The simplest example will be a $\ell=5$ vortex entering into a $C_3$ symmetric medium, $k_0=2$. We show 
in Fig.~\ref{f3} the phase and the dual of the flux field, i.e. that which is perpendicular at every point to $\vec\nabla\varphi$,  after evolving in a $C_3$ medium. From the analysis of the singularities we observe  one singularity of charge $m=-1$ in the center and two clusters of vortices, that we name {\it "waves"}, emanating 
from the center. Each one of these waves is formed by a cluster made of three vortices of charge +1. Therefore, 
$|k_0|$ stands for the number of waves emanating from the center.

Note that phase dynamics occurs close to the symmetry axis, i.e., in the region where the field "feels" the presence of the potential.  On the contrary, the phase is not modified by the existence of the potential  in points located far away from this region, since the potential quickly vanishes at long distances. Consequently, if we consider closed curves $\zeta$  surrounding the symmetry axis sufficiently far away from the region where the potential is significantly different from zero, the topological charge 
$v=\frac{1}{2\pi}\oint_{\zeta}\nabla\varphi\cdot d\mathbf{l}$, is equal to $\ell$. If this curve coincides with the boundary of $\mathbb{R}^2$, Eq.~(\ref{Eq1}) can be re-interpreted as a {\it total topological charge conservation law}. 

This is similar to the case of the total electrical charge, whose conservation can be derived from the 
$U(1)$ local gauge invariance of the SM lagrangian. 
The similarity between topological and electrical charges was introduced in the analysis of the XY model for interacting spin systems in two 
spatial dimensions~\cite{Kos73}. Within this model the total energy of a set of vortices with total topological 
charge equal to zero is the same as the one obtained for a diluted 2D Coulomb gas, therefore identifying
a set of vortices and antivortices with a neutral plasma formed by particles and antiparticles. In an analogous way, Eq.~(\ref{Eq1}) states that the the decay of an initial high-charged particle in a set of charged particles conserves the total charge. 

\begin{figure}[t]
\centering
\begin{tabular}{cc}
(a)&(b)\tabularnewline
\includegraphics[height=3cm]{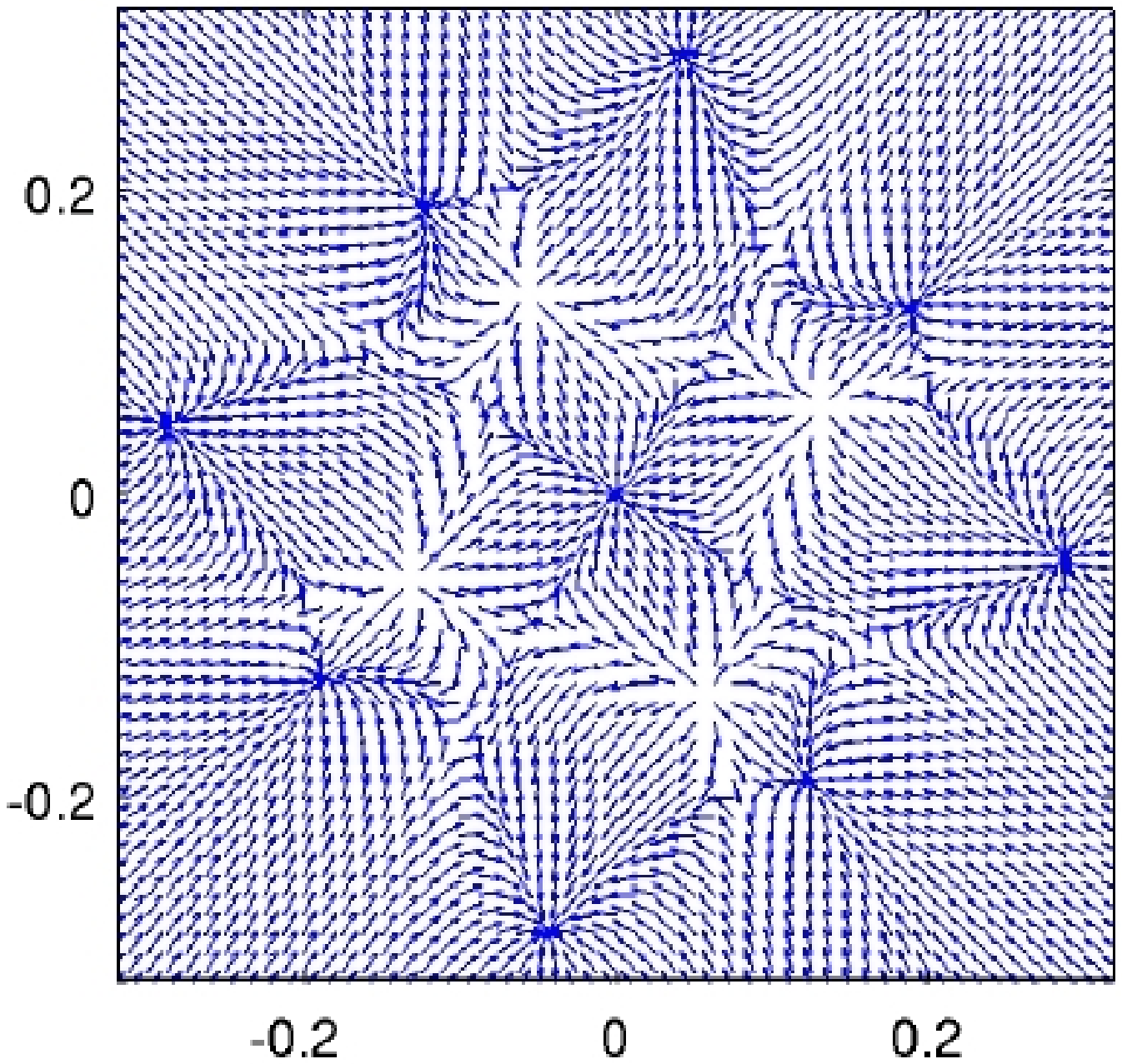}&
\includegraphics[height=3.25cm,width=3.25cm]{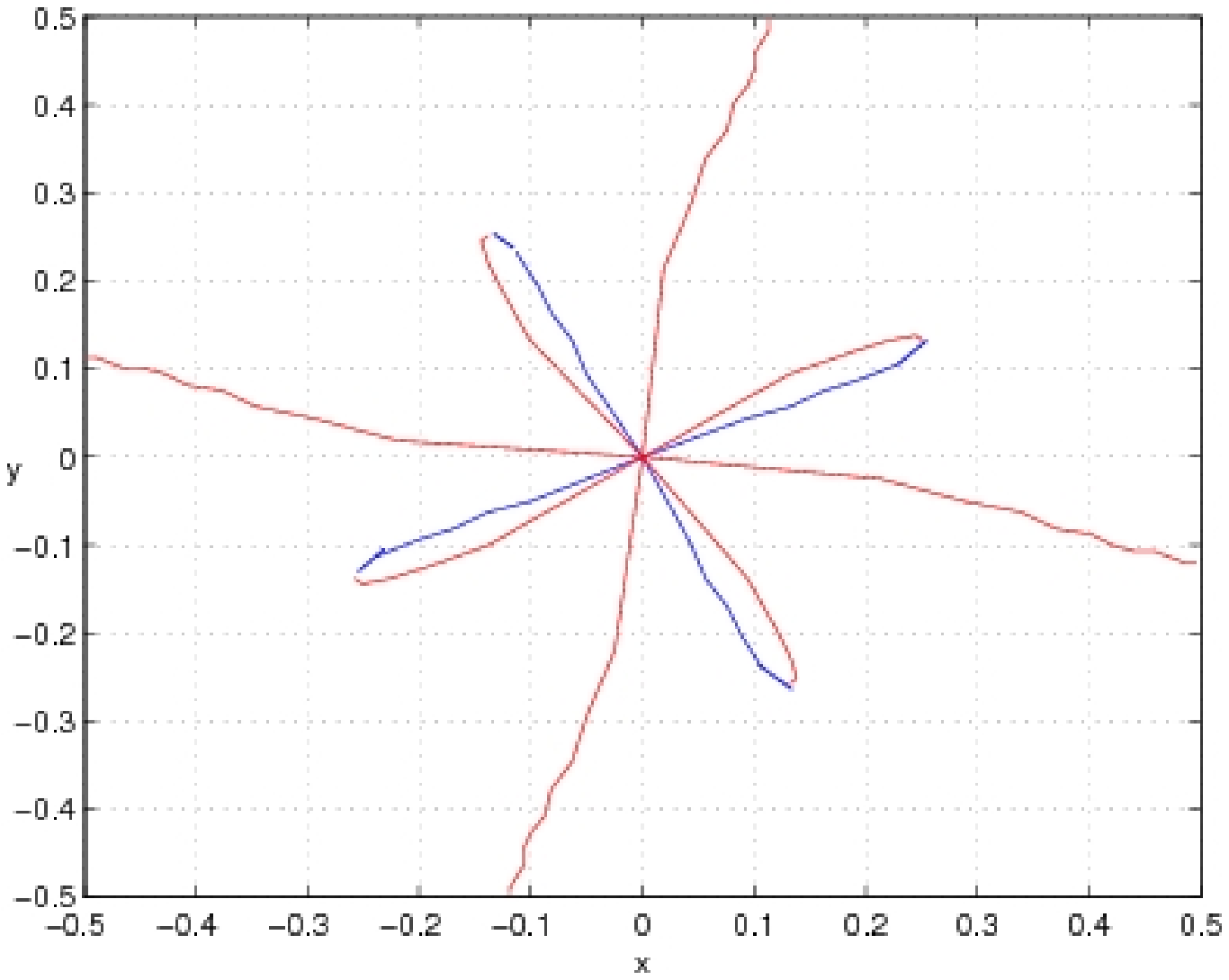}\tabularnewline
\end{tabular}
\caption{(a) Dual of the flux field for an initial vortex with angular momentum $\ell=5$ after 
evolving to $z=0.04$ into a $C_4$ symmetric medium tailored as explained in the text, and (b) vortices trajectories after evolving to $z=0.1$.}
\label{f4}
\end{figure}

Next we will see that this equation is still satisfied if vortex-antivortex pairs are generated.
According to Eq~(\ref{Eq4})  the coupling among the different field angular components is given by the potential terms $\overline{V}_k$. Therefore, by modifying the potential components one can manipulate the coupling between the different angular components, and hence the clusterization pattern. In the aforementioned examples, the potential components are decreasing. The linear potential $V_L$ can be engineered to modify this property, in such a way that the clusterization pattern can be controlled.
Let us define a potential where $\overline{V}_2>\overline{V}_1$ with a form $V_{C_n}(r,\theta)=g(r)\left[V_0+V_1\cos(n\theta)+V_2\cos(2n\theta)\right]$.  Within BEC's this potential would be obtained by using laser-interference based optical traps. 
In optics, it can be obtained by properly controlling the refractive index profile or by inducing it by means of interfering beams in a photorefractive medium. 
We show in Fig.~\ref{f4}(a) the dual of the flux  field of a vortex with angular 
momentum $\ell=5$ after evolving to $z=0.04$ into a $C_4$ symmetric medium characterized 
by such a potential. According to the previous results, one may expect a static vortex in the center with 
charge $+1$ and a cluster of four $+1$ charged vortices emanating away from it. However, 
the dynamics obtained is more involved. One static vortex in 
the center with charge $+1$ and three waves of vortices have been generated. 
The first two waves are made of four vortices with charge $+1$ each, while the second one is made of four $-1$  charged vortices (antivortices). Therefore, although the conservation rule holds independently of the choice of the potential, the wave dynamics 
is potential-dependent. We plot in Fig.~\ref{f4}(b) the vortices trajectories after evolving to $z=0.1$ (see EPAPS document No. [EPAPS-02.gif] for a 3D animation of the process). It can be observed how if the system is 
allowed to evolve beyond a critical distance the four antivortices annihilate with four vortices in the same way particles 
does, radiating their energy away, recovering in this way the picture naively expected. 
This phenomena, well know in particle physics, is usually referred 
to as pair production/annihilation. Quantum number conservation allows the creation of pairs of particles-antiparticles 
providing that all the conserved quantum numbers of the produced pairs sum zero. Therefore, since according to
the Eq. (\ref{Eq1}) the total topological charge is conserved, the topological charge that goes into the waves ($k_0n$) 
can be distributed in any manner consistent with the discrete group symmetry, creating $|k_v|+|k_0|$ 
waves of $n$ vortices together with $|k_v|$ waves of $n$ antivortices.

 The previous results reinforce the 
particle$-$phase singularity picture. A topological charge conservation rule has been identified and analysed. This relation allows to describe the vortex decay pattern into a cluster of vortices and to fully specify their number and properties, i.e., $m$ and the number of vortex,  $|k_0|$, and vortex-antivortex, $|k_v|$, waves.
This process can be engineered by  choosing the initial angular momentum $\ell$, the symmetry of the discrete medium $C_n$, and by tailoring the experimental $V_{C_n}(r,\theta)$ 
potential. This allows us to define a control mechanism which  would allow experimentalist to generate  and manipulate any desired configuration of waves of singularities. Finally, since all the results holds for any vortex  embedded into a $C_n$ symmetric medium, it opens the possibility to address a wide variety of phenomena in several different 
physical systems.

This work was partially supported by Generalitat Valenciana (Contracts
No. ACOMP07/221 and No. APOSTD/2007/052), and by the Government of Spain (Contract No. FIS2005-01189 and TIN2006-12890).

\end{document}